\documentstyle[prl,aps]{revtex}

\begin{document}
\preprint{TAUP 2340-96}
\title{Applications of the Complex Geometric ``Phase'' for Meta-stable Systems}
\author{S. Massar}
\address{School of Physics and Astronomy,
Raymond and Beverly Sackler Faculty of Exact Sciences, \\
Tel-Aviv University, Tel-Aviv 69978, Israel.}
\date{}
\maketitle
\begin{abstract}
Garrison and Wright showed that upon undergoing 
cyclic quantum evolution a meta-stable state acquires both a
geometric phase and a geometric decay probability.
This is described by a complex geometric ``phase'' associated with
the cyclic evolution of two states and is closely related to the two
state formalism developed by Aharonov et al.. Applications of the 
complex geometric phase to 
 the Born--Oppenheimer approximation and the
Aharonov--Bohm effect are considerd. A simple experiment
based on the optical properties of absorbing birefringent
crystals is proposed.
\end{abstract}

In quantum mechanics a state is define up to an arbitrary
phase. However phase differences have physical meaning. Therefore if a
state evolves in such a way that $\vert \psi(T)\rangle = e^{i\alpha}
\vert \psi(0)\rangle$, the phase $e^{i\alpha}$ 
can be measured. The remarkable property
discovered by Berry in the case of adiabatic evolution\cite{Berry} 
and subsequently
generalized by Aharonov and Anandan to arbitrary evolution\cite{AA} is that the
phase $e^{i\alpha}$ contains not only a dynamical term but also a
purely geometric term, the Berry or AA phase, which depends only on
the geometry of the circuit described by $\vert \psi(t)\rangle$ in
Hilbert space between $t=0$ and $t=T$. It is the holonomy acquired by 
$\vert \psi(t)\rangle$ when it is parallel transported around the 
circuit\cite{Simon}. Explicitly it takes the form
\begin{equation}
\phi^{AA} = -i \ln \langle \psi(0) \vert \psi(T)\rangle
+ i \int_0^T\!dt\ \langle  \psi(t) \vert \partial_t \psi(t)\rangle
\label{AAp}
\end{equation}
where the second term is the subtraction of the dynamical phase.

Berry's and AA's phase have been well verified experimentally in
such diverse contexts as optics, NMR, molecular
physics\cite{CW&TC}--\cite{General}. They have also surfaced in a wide variety of
theoretical contexts\cite{General}.
Berry's phase has been generalized to cyclic evolution of degenerate
states\cite{WZ}
and to non adiabatic evolution\cite{Nonad}. 

A generalization which has
received little attention is Garrison and Wright's application to
meta-stable states\cite{GW}. Garrison and Wright considered a meta-stable
state which undergoes cyclic evolution. They showed that 
if the meta-stable system has not decayed
during the time $T$, then its state $\vert \psi(T)\rangle$ coincides
with the original state $\vert \psi(0)\rangle$ up to a phase, which
can as expected be decomposed into a dynamical and a geometrical
term. However a new feature arises in this case, namely the
probability for the meta-stable state not to decay during the cyclic evolution
can also be decomposed into a dynamical and a geometrical
factor. 
This is described by a complex geometric ``phase'', the real part of
which corresponds to the geometric phase and the imaginary part to the
geometric decay probability. The complex geometric phase is associated with
the cyclic evolution of two states, contrary to the AA phase which is
associated with the cyclic evolution of one state.

Recently Aharonov et al. have shown how to describe quantum
systems in terms of two states, one evolving towards the future and one
evolving towards the past\cite{AV2}\cite{AV3}\cite{AR}. 
In particular they showed that measured
quantities which in the conventional one state approach are real
become complex in the two state approach. It has also been shown that
measurements on meta-stable systems naturally fit into  the two state
formalism\cite{1234}. It is therefore natural that 
the geometric quantity associated to the cyclic evolution of 
metastable states is described by the cyclic evolution of two states
and that it is complex.

Garrison and Wright illustrated how the complex geometric phase
arises in the case of an excited atom in a cyclically varying laser
field. The purpose of this article is to further analyse the 
properties of the complex geometric phase and to 
 propose several contexts
of theoretical or experimental interest in
which the complex geometric phase arises. In
the Born-Oppenheimer  approximation in the presence of meta-stable
systems 
it can be realized as
a complex vector potential for the slow degrees of freedom. This could
have applications when calculating the lifetime and energy levels  of excited
electronic-vibrational-rotational spectra of molecules. In the
Aharonov-Bohm effect it gives rise to topological decay probabilities
in addition to the usual topological phase.  
Finally we suggest a simple experimmental scheme in which to verify
Garrison and Wright's complex geometric phase which is
based on the optical properties of absorbent birefringent crystals.

Let us first recall Garrison and Wright's derivation of the complex geometric phase.
The time evolution of a meta-stable system is
\begin{equation}
\vert \psi(t)\rangle =
e^{-i\int_0^t\!dt\ H_{eff}} \vert \psi(0)\rangle + {\mbox {decay
    products}}
\end{equation}
where $H_{eff}$ is the effective Hamiltonian which acts on the
meta-stable state\cite{WW}. The most interesting situation arises when
there are several coupled meta-stable states which are nearly
degenerate.  For a review of some physical systems in which this occurs,
see\cite{Appl}.
We shall denote the space of meta-stable states ${\cal
  H}$. 
One can then consider that $H_{eff}$ acts only in   ${\cal H}$.
The eigenvalues $\omega_i$ of $H_{eff}$ are complex (we shall suppose
them non degenerate). The left and
right eigenvectors of $H_{eff}$
($\langle \phi_i\vert \omega_i = \langle \phi_i\vert H_{eff}$
and $H_{eff} \vert \psi_i \rangle = \omega_i  \vert \psi_i \rangle $)
each form a nonorthogonal basis of ${\cal H}$.
They obey the mutual orthogonality condition
$
\langle \phi_i\vert \psi_j \rangle
= \delta_{i j} \langle \phi_i\vert \psi_i \rangle
$
which follows from the double equality
$ \langle \phi_i\vert H_{eff} \vert \psi_j \rangle
= \omega_i  \langle \phi_i \vert \psi_j \rangle
=\omega_j  \langle \phi_i \vert \psi_j \rangle
$. Using this orthogonality condition  we can express $H_{eff}$ as
\begin{equation}
H_{eff} = \sum_i \omega_i { \vert \psi_i \rangle \langle \phi_i\vert 
\over \langle \phi_i\vert \psi_i \rangle}
\end{equation}

Let us now suppose that $H_{eff}$ is slowly changing with time.
The
amplitude for the meta-stable state not to decay is solution of the
effective Schr\"odinger equation $i \partial_t \vert \psi \rangle
= H_{eff}(t)   \vert \psi \rangle$. We can 
decompose  $\vert \psi \rangle$ into the
basis of instantaneous eigenkets
$\vert \psi \rangle = \sum_i a_i(t) e^{-i \int^t_0\! dt\ \omega_i(t)}
\vert \psi_i (t)\rangle$. Inserting this expression 
into the effective Schr\"odinger
equation and taking the scalar product with $\langle \phi_i\vert$ yields
\begin{equation}
\partial_t a_i + a_i {\langle \phi_i \vert \partial_t \psi_i\rangle
\over \langle \phi_i \vert\psi_i\rangle }
= - \sum_{j\neq i} a_j e^{-i \int^t_0\!dt\ (\omega_j - \omega_i)}
{\langle \phi_i \vert \partial_t \psi_j\rangle
\over \langle \phi_i \vert\psi_i\rangle }
\label{eqai}
\end{equation}
For sufficiently slowly varying Hamiltonian the right hand side of
this equation can be neglected (see discussion below).
Thus if the initial state
$\vert \psi(0) \rangle $ coincides with the eigenstate 
$\vert \psi_i \rangle $, in the adiabatic limit the solution of
the effective Schr\"odinger equation is
\begin{equation}
\vert \psi(t)\rangle 
=  e^{-i \int^t\!dt\ \omega_i}
 e^{-\int^t\!dt\ {\langle \phi_i \vert \partial_t \psi_i\rangle
\over \langle \phi_i \vert\psi_i\rangle }} \vert \psi_i (t) \rangle
\end{equation}
When $H_{eff}$ has evolved cyclically the second factor is a purely
geometric quantity associated with the cyclic evolution of the
meta-stable state:
\begin{eqnarray}
\phi_{GW}^i &=&
i\int_0^T\!dt\ {\langle \phi_i \vert \partial_t \psi_i\rangle
\over \langle \phi_i \vert\psi_i\rangle }\nonumber\\
&=&
{i \over 2}
\int_0^T\!dt\  {\langle \phi_i \vert \partial_t \psi_i\rangle
 - \langle \partial_t \phi_i \vert \psi_i\rangle
\over \langle \phi_i \vert\psi_i\rangle }
+ \partial_t  \ln  {\langle \phi_i(t)\vert \psi_i(t)\rangle}\label{phiC}
\end{eqnarray}
The conditions  $\vert \psi_i(T)\rangle = \vert \psi_i(0)\rangle$
and $\langle \phi_i(T)\vert = \langle \phi_i(0)\vert$ imply that
the boundary term in the second equality vanishes.
Further manipulation yields  an expression for 
$\phi_{GW}$ which is independent of the choice of
phase
in the definition of $\langle \phi_i (t)\vert$ and
$\vert \psi^i  (t)\rangle$:
\begin{eqnarray}
\phi^i_{GW} &=&
- {i \over 2} \ln {\langle \psi_i(0)\vert \psi_i(T)\rangle
\over \langle \phi_i(T)\vert \phi_i(0)\rangle}
+ {i \over 2}
\int_0^T\!dt\  {\langle \phi_i \vert \partial_t \psi_i\rangle
 - \langle \partial_t \phi_i \vert \psi_i\rangle
\over \langle \phi_i \vert\psi_i\rangle }
\label{phiC2}
\end{eqnarray}
This expression exhibits the symmetric role of $\langle \phi_i\vert$
and $ \vert \psi_i\rangle$ and shows  explicitly that $\phi_i$ is associated
only with the cyclic evolution of the two states  
and not with the structure of the
effective Hamiltonian. 
Thus eq.  (\ref{phiC2} is the generalization of geometric phases to
systems described by two states\cite{AV2}. When the two
states coincide one recovers the AA phase eq. (\ref{AAp}).

The real  part of $\phi^i_{GW} $ is the geometric
phase and the imaginary part yields the geometric decay probability. 
Their is however an important difference between the real and imaginary part
of $\phi_{GW}$. Indeed the geometric phase 
$(=Re(\phi_{GW}))$ is defined only for 
cyclic evolution. On the other hand the 
geometric decay probability $(=Im(\phi_{GW}))$ 
is defined even for non cyclic evolution. This is 
because the probability to have decayed is defined at all times.
One verifies that the imaginary part of eq. (\ref{phiC2}) is
a geometric quantity even for non cyclic evolution since it is 
independent of the choice of phase for $\langle \phi_i\vert$
and $ \vert \psi_i\rangle$ and it is independent of the
reparametrisation
of the path.
In this article we shall mostly consider the case of cyclic evolution
for which the analogy with Berry's phase is closest..

For cyclic evolution,
the integral in eq. (\ref{phiC2}) is around a contour, denoted ${\cal
  C}$, in the product, ${\cal H}\otimes {\cal
  H}$, of the space of meta-stable states. 
It can be reexpressed as the integral of a complex
two form over any surface $\partial{\cal C}$ with ${\cal C}$ as boundary
\begin{eqnarray}
\phi_{GW}&=& \int\!\int_{\partial{\cal C}}
\! dx^a\wedge dx^b B_{ab}\label{TwoF}\\
B_{ab}&=&
{\langle \partial_b \phi\vert \partial_a \psi \rangle
\over
\langle \phi\vert \psi \rangle}
-
{\langle \partial_b \phi\vert \psi \rangle
\langle \phi\vert \partial_a \psi \rangle
\over
\langle \phi\vert \psi \rangle^2}
-(a \leftrightarrow b)\label{B}
\end{eqnarray}
Garrison and Wright illustrated this formula in a particular case
for which $\phi_{GW}$ could be interpreted as a
complex solid angle, in generalization of Berry's result.

Let us return to the condition of validity of the adiabatic
approximation in eq. (\ref{eqai}). This is not straightforward because
the right hand side of eq. (\ref{eqai}) contains exponentially growing
terms (since the $\omega_i$ are complex). However it was shown in
\cite{HP} that in the case of hermitian Hamiltonians the amplitude for
non adiabatic transitions are exponentially small. Since the formal
expression for the non adiabatic transitions is the same in both
cases, the adiabatic approximation will be valid provided
$ {\rm Re} (\omega_i - \omega_j)$ is much larger than the other 
frequencies which appear in this problem. 

It is interesting to also consider
the opposite limit wherein the time $T$ it takes $H_{eff}$ to change
cyclically is much less than $\vert \omega_i - \omega_j \vert^{-1}$  
(but $T$ must nevertheless be long enough to ensure that the
concept of an effective Hamiltonian remains valid at all times). 
Then $H_{eff}$ can be reexpressed as $H_{eff} = \omega \sum_i {\vert
  \psi_i\rangle \langle \phi_i \vert \over \langle \phi_i \vert \psi_i
  \rangle }$. In this case 
the cyclic time evolution is given by the
operator:
\begin{eqnarray}
e^{-i \int_0^T\! dt\  H_{eff}} = e^{-i \omega t} \hat {\rm T} 
e^{i \int_0^T
  \!dt\ A(t)}\nonumber\\
A(t) =i \sum_{i , j} \vert \psi_i \rangle {\langle \phi_i\vert \partial_t
  \psi_j \rangle \over \langle \phi_i \vert \psi_i\rangle} \langle
\phi_j \vert
\end{eqnarray}
where $\hat {\rm T}$ is the time ordering operator. $A(t)$ is a
``non-hermitian
non-Abelian gauge potential'' which generalizes to
meta-stable states the non-Abelian gauge potential found by
Wilczek and Zee\cite{WZ} in the case of cyclic evolution of
degenerate states.

We now consider several applications of the complex geometric phase.
We first recall that the Berry phase was originally introduced in the
context of the Born Oppenheimer approximation where it appears as a
non trivial vector potential for the slow degrees of freedom\cite{MT}. 
In a similar way one
can consider the Born Oppenheimer approximation for a system composed
of rapid but meta-stable particles coupled to a slow system. 
The total Hamiltonian for such a system is
\begin{equation}
H_{eff} = {P^2\over 2M} + V(Q) + h_{eff}(q,Q)
\end{equation}
where $q$ is the fast degree of freedom and $Q$ is the slow degree of
freedom. Let $\vert \psi_i(q,Q)\rangle$ and $\langle \phi_i (q,Q)\vert$
be the instantaneous eigenstates of the rapid hamiltonian $h_{eff}(q,Q)$.
Postulating a wave function of the form $\vert \Psi \rangle = \vert
\chi_i(Q)\rangle\vert \psi_i(q,Q)\rangle$, one obtains for $\vert
\chi_i\rangle$ the equation
\begin{equation}
\left[ {1\over 2M}
\left( P - A_i(Q)\right)^2  + {\cal V}_i(Q)\right] \vert\chi_i\rangle
= \Omega_i\vert\chi_i\rangle
\end{equation}
where $\Omega_i$ is the complex energy of the eigenstate and
\begin{eqnarray}
{\cal V}_i(Q) &=& V(Q) + \omega_i (Q) + {1\over 2M}\left(
{ \langle \phi_i \vert \partial_Q^2 \psi_i\rangle \over \langle \phi_i
\vert \psi_i\rangle}
- {\langle \phi_i \vert \partial_Q\psi_i\rangle^2 \over \langle \phi_i
\vert \psi_i\rangle^2}
\right)\nonumber\\ 
A_i(Q) &=& -i {\langle \phi_i \vert \partial_Q\psi_i\rangle\over \langle \phi_i
\vert \psi_i\rangle}
\end{eqnarray}
The complex vector potential $A_i(Q)$ which
arises in this case could have measurable effects on the lifetimes and
energy levels  of
excited
electronic-vibrational-rotational molecular states\cite{General}.

Among the most interesting problems which can be 
analysed using the concept of geometric phase are the 
Aharonov-Bohm\cite{AB} and Aharonov-Casher\cite{AC} effects.
In this case the phases are purely topological, ie. they depend only on 
the winding number of the trajectory. We shall show that in the presence
of metastable particles their is also a topological decay probability,
ie. a decay probability that depends only on the winding number of the 
trajectory. This new feature arrises because the AB or AC phase becomes
complex in the presence of decaying particles.

Let us recall the AC effect (for reasons that will be discussed below,
this would be easier to realise experimentally than the AB effect). It 
consists of a particle with a magnetic moment moving in the presence
of a
charged line.
The charged line is taken along the $z$ axis.
The hamiltonian for the particle is\cite{AC}
\begin{equation}
H={1\over 2 m}
({\bf p} + \mu_z {\bf  a} ( {\bf r} ))^2 + V({\bf r})
\label{Hmu} 
\end{equation}
where $\mu_z$ is the projection of the magnetic moment along the $z$
axis
and in the cyclindrically symmetric gauge ${\bf  a} ( {\bf r} )$ takes
the form
\begin{equation}
a_\theta = {\rho \theta \over 2 \pi}\quad \ ,\ \quad a_r=a_z=0
\label{a}
\end{equation}
where $\rho$ is the charge per unit length of the  charged line  and 
$\theta$ is the angle around $z$ axis in cyclindrical coordinates.
Consider the amplitude for the particle to go from $P_1 = (r_1,
\theta_1, z_1)$ to $P_2 = (r_2,
\theta_2, z_2)$ in time $t$. We express it  using the
Feynman path integral and  decompose the sum over paths  in terms of the winding 
number of the path:
\begin{eqnarray}
K(P_2,P_1,t) &=& \int {\cal D} (x) e^{i S(x,\mu)}\nonumber\\
&=&e^{i \mu_z \rho (\theta_2 - \theta_1)}
\sum_n K_n^0(P_2,P_1,t) e^{i \mu_z \rho n} 
\label{K}
\end{eqnarray}
where 
\begin{equation}
K_n^0 = \int_{paths\ with\ winding\ number=n} {\cal D} (x) 
e^{i S(x,\mu=0)}
\end{equation}
is the contribution of paths with winding number $n$ when $\mu=0$.
The factor $e^{i \mu \rho n}$ is the AC phase. It affects interferences 
between contributions with different winding number.

Let us now turn to the case when the particle with magnetic moment is 
metastable. Suppose that their are several nearly degenerate
metastable states. Then as before their is an effective hamiltonian 
$h_{eff}^{internal}$
which acts in the space of metastable states. The left and right eigenstates
of $h_{eff}^{internal}$
will be denoted $\langle \phi_i^{int}\vert \vert \psi_i^{int}\rangle$
and their eigenvalue $\omega_i$.
The essential point in the derivation of the complex AC phase is
is that if the particle is in the eigenstate 
$\vert \psi_i^{int}\rangle$ and if the coupling of the magnetic 
moment to external systems is sufficiently weak and slowly varying,
then the effective value of 
the magnetic moment of the particle is(the proof is given below, see
also \cite{1234})
\begin{equation}
{\bf \mu}^i = {\langle \phi_i^{int} \vert {\bf \mu} \vert \psi_i^{int} \rangle
  \over  \langle \phi^{int}_i \vert \psi_i^{int} \rangle}
\label{mui}
\end{equation}
This expression for ${\bf \mu}^i$ is nontrivial if the magnetic moment operator
$\mu$ does not commute with $h_{eff}^{internal}$ whereupon
${\bf \mu}^i$ can be complex.
If the conditions for ${\bf \mu}$ to be  effectilvely given by
eq. (\ref{mui}) are satisfied,
the amplitude for the particle to go from
$P_1$ to $P_2$ and not to decay can be expressed in analogy with
eq. (\ref{K}) as
\begin{eqnarray}
K(P_2,P_1,t,no\ decay) =
e^{i \mu_z^i \rho (\theta_2 - \theta_1)}
\sum_n K_n^0(P_2,P_1,t) e^{i \mu_z^i \rho n} 
\label{Kdecay}
\end{eqnarray}
As before their is a topological phase given by 
$e^{i {\rm Re}(\mu_z^i) \rho n}$ . Their is also a topological 
decay probability given by $e^{- {\rm Im}(\mu_z^i) \rho n}$
which depends only on the winding number of the path.
In addition their is an overall geometric decay probability
$e^{- {\rm Im}(\mu_z^i) \rho (\theta_2 - \theta_1)}$ which depends
on the angle between $P_1$ and $P_2$ but is independent of the details of 
the path between $P_1$ and $P_2$.

We now derive this result in a more rigorous fashion.
When the particle is metastable one must add to the Hamiltonian eq. 
(\ref{Hmu})
the effective hamiltonian governing the internal state of the particle
\begin{equation}
H_{eff}={1\over 2 m}
({\bf p} + \mu_z {\bf  a} ( {\bf r} ))^2 + V({\bf r})
+h_{eff}^{internal}
\label{Hmueff} 
\end{equation}
with ${\bf a}$ once more given by eq. (\ref{a}).
Let us postulate a solution of the effective Schr\"odinger equation of
the form $\vert \Psi_i\rangle =\chi_i( {\bf r})
\vert \psi_i^{int} \rangle e^{-i \omega_i t}$. Inserting this ansatz into the effective
Schr\"odinger equation
and taking the scalar product with $\langle \phi_i^{int} \vert$ yields
\begin{equation}
-i \partial_t \chi_i =\left(
{1\over 2 m}
({\bf p} -  \mu _z^i{\bf  a})^2 +  V({\bf r}) \right)\chi_i 
\end{equation}
where we have neglected a term of the form   
${e^2 a^2 \over 4 m^2}({\langle \phi_i \vert \mu_z^2 \vert \psi_i
  \rangle
\over \langle \phi_i \vert \psi_i
  \rangle } - \mu_z^{i2} )$.
One can further verify that non adiabatic transitions are
controled by terms of the form 
${\langle \phi_j\vert \mu_z \vert \psi_i \rangle \over 
\langle \phi_i\vert \psi_i \rangle } {{\bf a} .  {\bf p} \over m} e^{-i (\omega_i - \omega_j)t}$
and 
${\langle \phi_j\vert \mu_z^2 \vert \psi_i \rangle \over 
\langle \phi_i\vert \psi_i \rangle } {a^2 \over 2m} e^{-i (\omega_i - \omega_j)t}$.
All these terms can be neglected when the particle is sufficiently far from the 
line of charge (so that ${\bf a}$ is small) and is moving sufficiently
slowly (so that ${\bf p}$ is small). 
It is then legitimate to replace ${\bf \mu}$ by its effective value
${\bf \mu}^i$. 
The solution for $\chi_i$ is then (in the WKB approximation)
\begin{equation}
\chi_i = e^{- i E_i t} e^{i \int^{\bf r} ({\bf p} +  \mu _z^i{\bf
    a}). d{\bf r}}
\end{equation}
and one obtains a similar expression for the ket $\langle
  \Phi_i\vert$ solution of the effective
Schr\"odinger equation $i \partial_t \langle
  \Phi_i\vert = \langle
  \Phi_i\vert H_{eff}$ .  To make connection with the geometric phase,
consider the case of cyclic evolution and insert $\langle
  \Phi_i\vert$ and $\vert \Psi_i \rangle$ into eq. (\ref{phiC2})
to yield
\begin{equation}
\phi_{GW} =
\oint  ({\bf p} + \mu_z^i{\bf a}). d{\bf r}= n \mu_z^i \rho + \oint  {\bf p}. d{\bf r}
\end{equation}
where $n$ is the winding number of the trajectory. 

Several remarks are in order. 

-First recall that whereas the geometric
phase is defined only for cyclic evolution, the geometric decay
probability is not. However the decay probability is
topological even for non cyclic evolution since it depends only on
the end points of the trajectory and the winding number, not on the
details of the trajectory. 

-Second note that  ${\bf a}$ was taken in the
cyllindrically symmetric gauge. 
As shown in \cite{gauge} this is the simplest gauge to 
use when the magnetic moment operator does not commute with
the full Hamiltonian. The analysis in other gauges are possible
but more complicated and necessitate the introduction of  non 
conventional commutation relations.

-Third, in order to obtain topological decay probabilties 
in the AB effect it would be necessary to take the solenoid to
consist of one metastable particle. This is probably very difficult to 
realise experimentally. (It is obviously impossible to have an effective 
complex electric  charge since charge is a conserved quantity).

We now consider how $\phi_{GW}$ could be measured
in optical experiments. We recall that one of the simplest ways to measure
Berry's phase is to pass polarized light through a coiled optical
fiber. The cyclic change in propagation direction of the light implies
a cyclic change in polarization direction and hence a geometric phase
which can be materialized as a rotation of the direction of
polarization of the light after exiting from the fiber\cite{CW&TC}. A
simple generalization of this scheme would be to use an optical fiber
made up of absorbing dichroic material. In such mediums the
different polarizations are absorbed at different rates and therefore
the polarization eigenstates are in general non
orthogonal\cite{dichroic}. 
The coils of the fiber will then give rise to a
cyclic change in polarization direction and hence to a
geometric phase and a geometric attenuation of the beam. 

An alternative approach would be to change cyclically the composition
of the absorbing dichroic material through which the light is passing. For
instance consider a beam of polarized light which passes through
vacuum and two different 
absorbing dichroic crystals $A$ and $B$ (note that $B$ could consist of the
same crystal  as $A$ but rotated through an angle $\theta$). If the
beam passes successively through vacuum, $A$, vacuum, $B$,
vacuum then $\phi_{GW}=0$. However if the sequence is  vacuum, $A$, $B$,
vacuum then the polarization eigenstates describe a non trivial
circuit and $\phi_{GW}\neq 0$. A comparison of the two sequences allows
$\phi_{GW}$ to be measured.
In
this experiment the different interfaces play a similar role to the
successive reflections in \cite{CAGJW}.

I would like to thank Y. Aharonov, S. Popescu and L. Vaidman for very helpful
discussions and comments about this work.



\begin{thebibliography}{999}

\bibitem{Berry} M. V. Berry, Proc. R. Soc. A {\bf 392} (1984) 45

\bibitem{AA} Y. Aharonov and J. Anandan, Phys. Rev. Lett. {\bf 58}
  (1987) 1593

\bibitem{Simon} B. Simon, Phys. Rev. Lett. {\bf 51}
  (1983) 2167

\bibitem{CW&TC} R. Y. Ciao and Y. S. Wu, Phys. Rev. Lett. {\bf 57}
  (1986) 933,  A. Tomita and R. Y. Ciao, Phys. Rev. Lett. {\bf 57}
  (1986) 937

\bibitem{CAGJW} R. Bhandari and J. Samuel, Phys. Rev. Lett. {\bf 60}
  (1988) 1211; 
R. Y. Ciao, A. Antaramian, K. M. Ganga, H. Jiao and 
R. Wilkinson, Phys. Rev. Lett. {\bf 60}
  (1988) 1214; M. Kitano, T. Yabuzaki and T. Ogawa, Phys. Rev. Lett. {\bf 58}
  (1987) 523

\bibitem{RMN}D. Suter, K. T. Mueller and A. Pines,  
Phys. Rev. Lett. {\bf 60}
  (1988) 1218



\bibitem{General} ``Geometric Phases in Physics'' edited by
 A. Shapere and F. Wilczek, World Scientific, Singapore (1988); 
R. Jackiw, Comments on Atomic an Molecular Physics
  {\bf 21} (1988) 71 and references therein

\bibitem{WZ} F. Wilczek and A. Zee, Phys. Rev. Lett. {\bf 52}
  (1984) 2111


\bibitem{Nonad} M. V. Berry, Proc. R. Soc. A {\bf 414} (1987) 31;
R. Jackiw, Int. J. of Mod. Phys. A {\bf 3} (1988) 285

\bibitem{GW} J. C. Garrison and E. M. Wright, Phys. Lett. A {\bf 128}
  (1988) 177

\bibitem{AV2}  Y.Aharonov and L. Vaidman, {\it Phys.  Rev.}   
{\bf A 41} (1990) 11.

\bibitem{AV3}  Y.Aharonov and L. Vaidman, {\it J. Phys.} {\bf A 24} (1991) 2315.

\bibitem{AR}  Y.Aharonov and B. Reznik, {\it Phys.  Rev.}   {\bf A} to be published.


\bibitem{1234}Y. Aharonov, S. Massar, S. Popescu, J. Tollaksen and
L. Vaidman, ``Adiabatic Measurements on Metastable Systems'', 
quant-ph/9602011

 \bibitem{WW} V.F. Weisskopf and E.P. Wigner, 
{\it Z. Physics} {\bf 63}, 54 (1930), {\bf 65}, 18 (1930);

\bibitem{Appl} J. Bernstein, in ``Cargese Lectures in Physics'' edited
  by M. Levy, Gordon and Breach (1967)

\bibitem{HP} J. T. Hwang and P. Pechukas, J. Chem. Phys. {\bf 67} (177) 4640



\bibitem{MT} C. A. Mead and D. G. Truhlar, J. Chem. Phys. {\bf 70}
  (1979) 2284

\bibitem{AB} Y. Aharonov and D. Bohm, Phys. Rev. {\bf 115} (1959) 485

\bibitem{AC} Y. Aharonov and A. Casher, Phys. Rev. Lett. {\bf 53}
  (1984) 319

\bibitem{gauge} Y. Aharonov and J. Anandan, Phys. Lett. A {\bf 160}
  (1991) 493

\bibitem{dichroic} G. N. Ramachandran and S. Ramaseshan, Handbuch der
  Physik xxv (1961) 1







\end{thebibliography}
\end{document}